\documentclass{JHEP3}

\usepackage{amsmath,amssymb}
\usepackage{graphicx}

\newcommand{\mcb}{\mathcal{B}}
\newcommand{\mcl}{\mathcal{L}}
\newcommand{\mcs}{\mathcal{S}}
\newcommand{\mcv}{\mathcal{V}}
\newcommand{\evbg}[1]{\left. #1 \right\vert_\text{bg}}

\title{Stability of Scalar Fields in Warped Extra Dimensions}

\date{\today}

\preprint{NIKHEF/2010-017}

\author{S.~Mert Aybat\\
Nikhef Theory Group, Science Park 105, 1098 XG Amsterdam, The Netherlands\\
\email{maybat@nikhef.nl}}

\author{Damien P. George\\
Nikhef Theory Group, Science Park 105, 1098 XG Amsterdam, The Netherlands\\
\email{dpgeorge@nikhef.nl}}

\abstract{
This work sets up a general theoretical framework to study stability
of models with a warped extra dimension where $N$ scalar fields couple
minimally to gravity.  Our analysis encompasses Randall-Sundrum
models with branes and bulk scalars, and general domain-wall models.
We derive the Schr\"odinger equation governing the spin-0 spectrum
of perturbations of such a system.  This result is specialized to
potentials generated using fake supergravity, and we show that models
without branes are free of tachyonic modes.  Turning to the existence
of zero modes, we prove a criterion which relates the number of 
normalizable zero modes to the parities of the scalar fields.  
Constructions with definite parity and only odd scalars are 
shown to be free of zero modes and are hence perturbatively stable.
We give two explicit examples of domain-wall models with a soft wall,
one which admits a zero mode and one which does not.  The latter is
an example of a model that stabilizes a compact extra dimension using
only bulk scalars and does not require dynamical branes. 
}

\keywords{Field Theories in Higher Dimensions, Supergravity Models}
\begin{document}

\section{Introduction}\label{sec:intro}
Extra dimensions are a plausible extension to the Standard Model (SM),
and a vast array of specific realizations of models with extra dimensions exist.  String theory
provides a lot of the motivation to include extra dimensions, and
there are many string-inspired phenomenological models, for example 
Refs.~\cite{Antoniadis:1990ew, ArkaniHamed:1998rs, Antoniadis:1998ig,
Randall:1999ee, Randall:1999vf},
as well as purely field-theoretic constructions, for example Refs.~\cite{Rubakov:1983bb, Kehagias:2000au,Davies:2007xr}.

Much work has been devoted to the study of the type I Randall-Sundrum
(RS) model~\cite{Randall:1999ee}, where the electroweak hierarchy
can be naturally generated by the warping of spacetime in the
extra dimension.  Such warping is a generic feature of models with
an extra dimension: the warp factor of the background metric is
sensitive to \emph{changes} in the energy density as one moves along
the extra dimension, and so non-trivial background sources will
generically induce a warped metric.  In the simplest 5D set-up, the
line element of a warped space is
\begin{equation}
ds^2 = e^{-2\sigma(y)} \eta_{\mu\nu} dx^\mu dx^\nu + dy^2 \:,
\end{equation}
where $y$ is the extra dimension and $\sigma(y)$ is the warp factor.  Scalar fields with profile
$\phi_i(y)$ and fundamental branes with tension $\lambda_\alpha$
located at $y_\alpha$ will provide a source of energy density and
act to drive the warp factor:
\begin{equation}
\label{eq:warpbehaviour}
\sigma''(y) \propto \sum_i \phi_i'(y)^2
    + \sum_\alpha \lambda_\alpha \delta(y-y_\alpha) \:.
\end{equation}
Here, $i$ labels the scalars and $\alpha$ labels the branes.  Note
that scalars can only provide `positive warping', and only if they
have a non-trivial profile in the extra dimension.  Branes can
provide both `positive warping' and `negative warping' since
$\lambda_\alpha$ can be of either sign.

In the original RS set-up, the extra dimension was compactified
on a circle with $y$ identified with $-y$.  The two branes at the
orbifold fixed points were required to have tensions of equal
magnitude but opposite sign in order for the warp factor $\sigma$
to join up consistently; the positive and negative warping needed
to balance.\footnote{The model also required a bulk cosmological
constant to keep $\sigma'(y)$ constant between the two branes.
This constant needed to be fine-tuned against the tension of the
branes, a generic feature of warped extra dimensions with
4D Poincar\'e slices.}
The type II RS model~\cite{Randall:1999vf} had the extra dimension
infinite in size, and required only a single brane of positive
tension at the origin.  The warp factor is driven to $\pm\infty$
at large distances from the origin, localizing 4D gravity to the
brane.  Since this model contains only a positive tension brane,
one can replace it by a scalar field with a suitable $y$-profile.
Such domain-wall models also localize 4D gravity~\cite{Csaki:2000fc}
and form the basis of a large area of research, see for example Refs.~\cite{ Kehagias:2000au,Davies:2007xr}.

Recently there has been interest in a new type of RS-like warped
spacetime: a compact spacetime where the negative tension brane is
replaced with a physical singularity.  These soft-wall models were
originally designed to yield linear Regge trajectories in the
context of the AdS/CFT correspondence~\cite{Karch:2006pv}, but
have since been the basis of actual models beyond the Standard
Model (BSM)~\cite{Batell:2008zm, Falkowski:2008fz,Batell:2008me, Delgado:2009xb,MertAybat:2009mk,Gherghetta:2009qs,Cabrer:2009we,vonGersdorff:2010ht},
and also provide a holographic dual description of unparticle models~\cite{Cacciapaglia:2008ns, Falkowski:2008yr}.
Our ultimate aim is take the soft-wall model and go one step further by
removing the final, positive-tension brane at the origin, and
replacing it with a suitable scalar field profile. 
The model will
then describe a compact extra dimension without the use of
fundamental dynamical branes, in other words there are no branes with tension and localized potential terms in the Lagrangian. This is interesting because it is a purely field 
theoretical construction and requires no appeal to string theory.
\footnote{The examples we will consider in the following sections, however, will have orbifold fixed points at the origin. These are properties of the geometry and one does not need to appeal to string theory to construct them.}

Such a step is not as straightforward as it may first seem, as
it is crucial that one ensures the stability of the scalar field
configuration.  Part of this involves stabilizing the size of the
compact extra dimension.  For the RS model, one can utilize the
Goldberger-Wise mechanism~\cite{Goldberger:1999uk, Csaki:2000zn}
where a bulk scalar field stabilizes the set-up due to localized
potentials on each brane.  Here, the presence of branes and the
ability to have a 4D potential for the 5D scalar localized to the
brane is necessary for stabilization to work.  Soft-wall models
can also be stabilized in a similar way~\cite{Cabrer:2009we};
here, one needs a 4D potential localized to the brane at the
origin.  In order to build a soft-wall model without a
brane, we shall need an alternative to the Goldberger-Wise
mechanism.  It is the aim of the present paper to show that, with
the correct type of scalar background, one does not need any
additional stabilization mechanism.

The techniques developed in this paper have wider application than
just soft-wall models without a brane.  The problem of stability
of non-trivial scalar backgrounds is an import one, and in the
case of a single scalar without gravity there exists a
constructive way of finding the lowest energy, stable
solution~\cite{Toharia:2007xe, Toharia:2007xf}.  For the case with
gravity, things are not so simple, but some attempts have been
made~\cite{Kobayashi:2001jd,Toharia:2008ug, Toharia:2010ex}.  The problem is
difficult because, as discussed above, non-trivial scalar profiles
always induce warping of the metric, and, furthermore, modes of
the scalars mix with spin-0 degrees of freedom in the metric.  In
the literature to date, all successful analyses have relied on the superpotential approach~\cite{DeWolfe:1999cp} also known as the fake supergravity approach~\cite{Freedman:2003ax}.  Fake supergravity 
requires one to choose a scalar potential such that it is
invariant under supergravity transformations, even though the
whole theory itself is not locally supersymmetric.  In practice
this is very simple to achieve: one just needs to generate the
full potential from a more primitive superpotential.  Using this
fake supergravity approach, one can easily solve the background
equations of motion, including Einstein's equations. It has been shown, using such a construction, that models with an arbitrary number of scalars are free of tachyonic spin-2 modes~\cite{DeWolfe:1999cp}. For a single scalar field, it has also been shown that this construction leads to models free of tachyonic modes in the spin-0 sector~\cite{DeWolfe:2000xi}. In this paper we extend this work to the case of an arbitrary number of background scalar fields. We find that the same result holds: there are no tachyonic modes in the spin-0 sector. Restricting ourselves to orbifold spaces that are symmetric under parity, we also show that set-ups where the scalar profiles are all odd do not contain spin-0 zero modes.

The paper is organized as follows.  In Section~\ref{sec:gravity_free}
we discuss the gravity-free case as a warm-up and introduction to
the superpotential approach.  We show that a scalar field configuration
has a mode spectrum which is always free of tachyons, and which always
contains a zero mode of translation.  In Section~\ref{sec:background}
we describe the model with gravity and the background solutions in the
fake supergravity approach.  The spectrum of spin-2 perturbations
in this background are derived in Section~\ref{sec:spin2pert}.  The
usual result of a massless 4D graviton is reproduced.
Section~\ref{sec:spin0pert} contains the main results of this paper,
and shows that, in the fake supergravity approach with an arbitrary
number of background scalars, there are no tachyonic spin-0
perturbations.
This analysis is valid for an extra dimension of general topology.
In Section~\ref{sec:specificn} we specialize to orbifold constructions 
with definite parity and discuss a criterion for determining 
the existence of spin-0 zero modes.  We then analyze two
specific models: one which supports such a zero mode, and one which
does not.  The latter is an example of a model with a stabilized
compact extra dimension that does not require any dynamical branes.  We 
conclude in Section~\ref{sec:conclusions}.

\section{Gravity-free case}\label{sec:gravity_free}
As discussed in the introduction, non-trivial scalar profiles always induce
a warped metric, and so realistic models must include gravity.
Nevertheless, the gravity-free case is interesting to
study before moving to the case with gravity, and serves also
to introduce our notation.  For one scalar field the general analysis has been completed in Refs.~\cite{Toharia:2007xe, Toharia:2007xf} and for two scalar fields, using the superpotential approach, in Ref.~\cite{Bazeia:2010yp}. In this section we show that, for $N$ scalar fields without gravity, the superpotential approach yields solutions
which are globally stable, for given boundary conditions, and
up to zero-mode translations of the configuration. 

The action is
\begin{equation}
\mcs = \int d^4x \, dy \left[
    -\frac{1}{2} \sum_i \partial^M \Phi_i \partial_M \Phi_i
    - V(\{\Phi_i\})
\right] \:,
\end{equation}
where the sum is over $1, \ldots, N$, and there are $N$ scalar
fields.  The potential $V$ is an arbitrary function of these $N$
scalars.  The equations of motion are
\begin{equation}
\label{eq:gf-eom}
\partial^M \partial_M \Phi_i - V_i(\{\Phi_i\}) = 0 \:.
\end{equation}
Here we place a subscript $i$ on $V$ to denote partial
differentiation with respect to the field $\Phi_i$.  This notation
is used heavily throughout the paper; in general, for a function
$X$ that depends on the scalar fields, we write
\begin{equation}\label{eq:deriv-defn}
X_{ij\dots k}(\{\Phi_i\}) \equiv \frac{\partial}{\partial \Phi_i}\frac{\partial}{\partial \Phi_j}\dots\frac{\partial}{\partial \Phi_k} X(\{\Phi_i\})\:.
\end{equation}

We are interested in finding static background solutions
for the scalars, configurations that depend only on the extra
dimension: $\Phi_i = \phi_i(y)$.  In what follows, `configuration'
refers to the set of $N$ scalar fields taking on the static
solutions given by $\phi_i(y)$.

We can restrict ourselves to potentials generated by a superpotential
$W(\{\Phi_i\})$ such that
\begin{equation}
\label{eq:gf-v}
V(\{\Phi_i\}) = \sum_i \frac{1}{2} \left[ W_i(\{\Phi_i\}) \right] ^2 \:.
\end{equation}
The subscript $i$ denotes partial differentiation with respect to
$\Phi_i$, as per Eq.~\eqref{eq:deriv-defn}.  With this particular
choice of potential, the Euler-Lagrange equations of motion,
Eq.~\eqref{eq:gf-eom}, are satisfied so long as the static
configuration $\phi_i(y)$ solve the first-order differential equation
\begin{equation}
\label{eq:gf-phiw}
\phi_i'(y) = W_i(\{\phi_i\}) \:.
\end{equation}
A prime denotes a $y$ derivative, and we are evaluating $W_i$
at $\phi_i(y)$.  Note that to show this
solves the equations of motion, and in many places throughout this
paper, we use the relation
\begin{equation}
\frac{dW}{dy} = \sum_i \frac{\partial W}{\partial \Phi_i} \frac{d\Phi_i}{dy} \:.
\end{equation}

It turns out that such background configurations are always globally stable,
up to zero-mode translations.  To prove this statement, we will look
at three things: 1) the 4D energy density of the configuration; 2)
perturbative modes; 3) the zero mode.

We compute the 4D energy density of the system by integrating
the 5D stress energy over $y$.  We shall assume that $V$ takes the
form given by Eq.~\eqref{eq:gf-v} and that the scalar fields are
functions of $y$ only, given by $\phi_i(y)$.  In particular we do
not require the scalar fields to satisfy Eq.~\eqref{eq:gf-phiw}.  We
obtain
\begin{align}
E &= \int T_{00} dy \nonumber\\
  &= \int \left[ \frac{1}{2} \sum_i (\phi_i')^2
    + \frac{1}{2} \sum_i \left[ W_i(\{\phi_i\}) \right]^2 \right] dy \nonumber \\
\label{eq:grav-free-energy}
  &= \int dW + \frac{1}{2} \sum_i \int \left[ \phi_i' - W_i(\{\phi_i\}) \right]^2 dy \:.
\end{align}
The first term in the last line here is just a surface term; it is
difference of $W$ evaluated at the boundaries of the extra
dimension, and depends only on the values of $\phi_i$ at these
boundaries.  Given a particular $W$, which completely determines
$V$, and choices for the boundary values of $\phi_i$, the total
energy density of the configuration is minimized precisely when Eq.~\eqref{eq:gf-phiw} is satisfied.  Hence, for given
boundary conditions, the superpotential approach yields
configurations which globally minimize the energy.

Since the configuration minimizes the energy, the solution must be
perturbatively stable, and we can demonstrate this explicitly.
This will be useful as a precursor to the case with gravity. Expanding in linear perturbations $\varphi_i$ about the background
\begin{equation}
\Phi_i(x^\mu, y) = \phi_i(y) + \varphi_i(x^\mu, y) \:,
\end{equation}
the equations of motion, Eq.~\eqref{eq:gf-eom}, reduce to
\begin{equation}
-\varphi_i''
    + \evbg{\left( W_{ik} W_{kj} + W_{ijk} W_k \right)} \varphi_j
    = \Box \varphi_i \:.
\end{equation}
Here, repeated indices are
to be summed over, and $\evbg{X}$ indicates that $X$ is a function
of $\Phi_i$ and is to be evaluated using the background solutions
$\phi_i(y)$; that is, $X(\{\phi_i\})$.
Performing a Fourier transform, $\Box\varphi_i\to E\varphi_i$, we
obtain a set of $N$, coupled, time-independent Schr\"odinger
equations.  We can write these Schr\"odinger equations as~\cite{Bazeia:2010yp}
\begin{equation}
\label{eq:gf-schro}
(\partial_y \delta_{ik} + \evbg{W_{ik}})(-\partial_y \delta_{kj} + \evbg{W_{kj}}) \varphi_j
    = E \varphi_i \:,
\end{equation}
and so, using the results of Appendix~\ref{sec:hamiltonian_app}, we see the
system admits solutions only with $E\ge0$. This is true so long as the
perturbations vanish at the boundaries of the extra dimension which is a
valid assumption.

Now let us consider the existence of the zero-mode
perturbations with $E=0$.  For such a mode to exist, Eq.~\eqref{eq:gf-schro}
implies that $\varphi_i' = W_{ij} \varphi_j$, which admits the
solution $\varphi_i = a W_i = a \phi_i'$, with $a$ a real,
non-zero constant.  This is the familiar result that the zero mode
of translation is the first derivative of the background configuration.
Note that when there are multiple background fields, the zero mode
is the mode where all $N$ perturbations $\varphi_i$ are excited
simultaneously with profiles proportional to $\phi_i'$.

As a quick example, consider $W = -\sqrt{\lambda/2}(\phi^3/3-v^2\phi)$,
which yields the familiar potential $V=(\lambda/4)(\phi^2-v^2)^2$.
Solutions are $\phi=\pm v$ and the well-known kink:
$\phi=v\tanh\left[v\sqrt{\lambda/2}(y-y_0)\right]$.  To get the anti-kink solution,
one needs to start with $-W$ instead.  This demonstrates the fact
that the superpotential encodes for both $V$ and the static solution,
and that different superpotentials can yield the same potential.
Also noteworthy is the fact that, for this choice of $V$, $\phi=0$
is a solution of the equation of motion but \emph{cannot} be obtained
from any superpotential.  This is because the $\phi=0$ solution is
unstable, and we have shown that the superpotential approach always
yields globally stable solutions.

\section{Warped Background Configuration}\label{sec:background}
We now come to the main topic of the paper and consider a general
5D theory with gravity coupled minimally to $N$ scalar fields,
including the possibility of fundamental brane terms.  The
corresponding action is given by
\begin{equation}
\mcs = \int d^4x \, dy
    \left[ \sqrt{-g} \left( M^3 R + \mcl_{\text{matter}} \right)
        - \sqrt{-g_4} \lambda
    \right] \:.
\end{equation}
We are using a $(-++++)$ signature and $M$ is the 5D Planck mass.
The scalar-matter Lagrangian and brane terms are given
respectively by
\begin{align}
\mcl_\text{matter} &= -\frac{1}{2} \sum_i g^{MN} \partial_M \Phi_i \partial_N \Phi_i - V(\{\Phi_i\}) \:,\\
\lambda &= \lambda(\{\Phi_i\}) = \sum_{\alpha} \lambda_\alpha(\{\Phi_i\}) \delta(y-y_\alpha) \:.
\end{align}
The sum over $i$ is from 1 to $N$, and $V(\{\Phi_i\})$ is the
potential that in general depends on all $N$ scalars.  The
subscript $\alpha$ indexes the branes, $y_\alpha$ are their
locations, and $\lambda_\alpha(\{\Phi_i\})$ the scalar potentials
localized to the branes, which includes the brane's tension.
Our aim is to study the general stability conditions for
non-trivial background configurations of this model.  In this
section we discuss background solutions, and then in the following two sections analyze spin-2 and spin-0
perturbations.

For 4D Poincar\'e slices, the background metric ansatz is
\begin{equation}
ds^2 = e^{-2\sigma(y)} \eta_{\mu\nu} dx^\mu dx^\nu + dy^2 \:,
\end{equation}
where $\mu,\nu$ index the 4D subspace.
Einstein's equations are
\begin{equation}
G_{MN} = \frac{1}{2M^3} T_{MN} \:.
\end{equation}

Assuming the scalar fields only depend on the extra dimension $y$,
and denoting such a background configuration by $\phi_i(y)$, the
Euler-Lagrange and Einstein's equations yield $N+2$ equations for
$N+1$ functions:
\begin{equation}
\label{background_eqs}
\begin{aligned}
&3 \sigma'' =
    \frac{1}{2M^3} \left( \sum_i \phi_i'^2 + \lambda(\{\phi_j\}) \right) \:, \\
&12 \sigma'^2 - 3 \sigma'' =
    -\frac{1}{2M^3} \left[ 2V\left(\{\phi_j\}\right) + \lambda\left(\{\phi_j\}\right) \right] \:, \\
&\phi_i'' - 4 \sigma' \phi_i'
    - V_i(\{\phi_j\})
    - \lambda_i(\{\phi_j\}) = 0 \:.
\end{aligned}
\end{equation}
The first two equations are in fact the sum and difference of the
$(\mu\nu)$ and $(55)$ components of Einstein's equations.  The
notation $V(\{\phi_i\})$ means that the potential is to be
evaluated with the background fields $\phi_i$.  Also, we place
a subscript $i$ on $V$ and $\lambda$ to denote partial
differentiation with respect to the field $\Phi_i$, as per
Eq.~\eqref{eq:deriv-defn}.
It is easy to show that one of these equations can be obtained using two others, therefore the equations are not all independent.

Solutions to the above equations can be obtained using the so-called fake supergravity approach. In this approach a superpotential $W$ is introduced so that
\begin{equation}
V(\{\Phi_i\})
    = \sum_i \frac{1}{2} \left[ W_i(\{\Phi_i\}) \right]^2
    - \frac{1}{3 M^3} \left[ W(\{\Phi_i\}) \right]^2 \:,
\end{equation}
where $W_{i}$ is obtained using Eq.~\eqref{eq:deriv-defn}.
It was shown in Ref.~\cite{DeWolfe:1999cp} that the system of equations, Eqs.~(\ref{background_eqs}), can be written in terms of the superpotential $W$ and solutions are given by
\begin{equation}
\label{w}
\begin{aligned}
\sigma'(y) &= \frac{1}{6 M^3} W(\{\phi_i\}) \:,\\
\phi_i'(y) &= W_i(\{\phi_i\})\:.
\end{aligned}
\end{equation}
However, what we are interested in here are not the solutions of the
system of equations. Instead, assuming solutions exist, we would like
to see if the system is stable or not.

When studying the gravity-free case in the previous section, we were
able to obtain the result that the superpotential approach yields
globally stable configurations, Eq~\eqref{eq:grav-free-energy}.
Unfortunately, for the case with gravity included the corresponding
analysis does not tell us much.  The 4D energy density is (recall that
we do not assume the relations in Eq~\eqref{w})
\begin{align}
E &= \int T_{00} dy \nonumber\\
  &= \int e^{-2\sigma} \left[ \frac{1}{2} \sum_i (\phi_i')^2
    + \frac{1}{2} \sum_i \left[ W_i(\{\phi_i\}) \right]^2
    - \frac{1}{3M^3} \left[ W(\{\phi_i\}) \right]^2 \right] dy \nonumber \\
  &= \int \left[ e^{-2\sigma}W(\{\phi_i\}) \right]' dy
    + \frac{1}{2} \sum_i \int e^{-2\sigma} \left[ \phi_i' - W_i(\{\phi_i\}) \right]^2 dy \nonumber\\
    &\qquad\qquad
    + 2 \int e^{-2\sigma} W(\{\phi_i\}) \left[ \sigma' - \frac{1}{6M^3} W(\{\phi_i\}) \right] dy
    \:.
\end{align}
The first term here is just a surface term, which generally vanishes
for a Randall-Sundrum-like configuration.  The second term is a
non-negative integral which is minimized when the scalar fields
obey the first order fake supergravity equation (as in the gravity-free
case).  The final term is the difficult term.  Although it vanishes
when $\sigma$ satisfies its first order equation, it is not obvious
that this minimizes $E$.  So we cannot conclude that the fake
supergravity approach yields a globally stable background configuration.

To proceed we shall study local, perturbative stability of a configuration
with gravity coupled to $N$ scalars.  The spin-2 and spin-0 fluctuations
of the metric and the scalar fields are treated separately in the following
two sections.  For the initial stages, our analysis will be for an arbitrary
scalar potential $V$.  Later on we will need to specialize to the fake
supergravity approach.

\section{Spin-2 Perturbations}\label{sec:spin2pert}
The general ansatz which takes into account both spin-0 and spin-2
perturbations is
\begin{equation}
\label{generalansatz}
\begin{aligned}
&ds^2 = e^{-2 \sigma(y)} \left[ (1-2F(x^\mu,y) \eta_{\mu\nu} + h_{\mu\nu}(x^\mu,y) \right] dx^\mu dx^\nu
    + \left[1+G(x^\mu,y)\right]^2 dy^2 \:,\\
&\Phi_i(x^\mu,y) = \phi_i(y) + \varphi_i(x^\mu,y) \:.
\end{aligned}
\end{equation}
Here, we have chosen to work in the axial gauge, $h_{\mu5}=0$, with
transverse traceless part $\partial^\mu h_{\mu\nu} = \eta^{\mu\nu} h_{\mu\nu}=0$. The $G_{ij}$ (spatial) components of Einstein's equations enforce
$G=2F$ which we take from now on. The Einstein tensor has the following non-zero components
\begin{equation}
\begin{aligned}
G_{\mu\nu} &=
    3 e^{-2\sigma} \eta_{\mu\nu}
        \left( 2\sigma'^2 - \sigma'' - F'' + 6\sigma'F' - 12\sigma'^2F + 6\sigma''F \right) \\
    &\qquad + e^{-2\sigma}
        \left( -\tfrac{1}{2} e^{2\sigma} \Box h_{\mu\nu}
            - \tfrac{1}{2} h_{\mu\nu}''
            + 2\sigma' h_{\mu\nu}'
            + 3(2\sigma'^2 - \sigma'') h_{\mu\nu}
        \right) \:,\\
G_{\mu5} &= 3\partial_\mu(F' - 2\sigma'F) \:,\\
G_{55} &= 6\sigma'^2 - 3 e^{-2\sigma} \Box F + 12\sigma'F' \:.
\end{aligned}
\end{equation}
The stress-energy tensor is
\begin{equation}
\begin{aligned}
T_{\mu\nu} &= e^{-2\sigma} \eta_{\mu\nu}
        \left[
            -\frac{1}{2}(1-6F) (\phi_i')^2
            - \evbg{(V + \lambda)}
            + \evbg{(2 V + 4 \lambda)} F
            - \evbg{(V_i + \lambda_i)} \varphi_i
            - \phi_i' \varphi_i'
        \right] \\
    &\qquad + e^{-2\sigma} h_{\mu\nu}
        \left[
            -\frac{1}{2} (\phi_i')^2
            - \evbg{(V + \lambda)}
            - \evbg{(V_i + \lambda_i)} \varphi_i
        \right] \:,\\
T_{\mu5} &= \partial_\mu(\phi_i' \varphi_i) \:,\\
T_{55} &= \frac{1}{2} (\phi_i')^2 - \evbg{V} - 4 \evbg{V} F - \evbg{V_i} \varphi_i + \phi_i' \varphi_i' \:.
\end{aligned}
\end{equation}
Here, repeated $i$ indices are to be summed over.
Taking the 4-trace of the $(\mu\nu)$ part of Einstein's equations 
shows that the spin-2 and spin-0 perturbations completely decouple
from one another. See Ref.~\cite{Csaki:2000zn} for a discussion of the gauge degrees of freedom and the decoupling of spin-2 and spin-0 sectors. The spin-2 perturbations are the easiest to
analyze, and we look at them first, returning to the spin-0
sector in the following section.  The Euler-Lagrange equations for
the $N$ scalar fields do not contain $h_{\mu\nu}$; these
equations are given in the next section.

Using the background equations, Eq.~\eqref{background_eqs}, the equation
for $h_{\mu\nu}$ is
\begin{equation}
-e^{2\sigma} \Box h_{\mu\nu}
    - h_{\mu\nu}''
    + 4\sigma'h_{\mu\nu}' = 0 \:.
\end{equation}
There is always a zero mode, $h_{\mu\nu}'=0$, which is
normalizable.  It is the well-known 4D massless graviton~\cite{Randall:1999vf}. In conformal coordinates $z$ defined by $dy=e^{-\sigma}\,dz$ with rescaled $h_{\mu\nu} = e^{3\sigma/2} \tilde{h}_{\mu\nu}$
we have
\begin{equation}
-\tilde{h}_{\mu\nu}''
    + \left( \frac{9}{4} \sigma'^2 - \frac{3}{2} \sigma'' \right)
    = \Box \tilde{h}_{\mu\nu} \:.
\end{equation}
This is a Schr\"odinger-like equation.  It can be rewritten in a self-adjoint form as
\begin{equation}
(\partial_z + S_h)(-\partial_z + S_h) \tilde{h}_{\mu\nu}
    = \Box \tilde{h}_{\mu\nu} \:,
\end{equation}
with
\begin{equation}
S_h = -\frac{3}{2}\sigma' \:.
\end{equation}
Appealing to Appendix~\ref{sec:hamiltonian_app}, we see that there are no tachyonic modes in the spin-2 sector.

\section{Spin-0 Perturbations}\label{sec:spin0pert}
In the previous section we have seen that the spin-2 sector decouples from the spin-0 sector, independent of the number of scalar fields involved. We now consider spin-0 perturbations around the background solutions
$\sigma(y)$ and $\phi_i(y)$. The metric ansatz is a restricted version of Eq.~\eqref{generalansatz} with $h_{\mu\nu}=0$~\cite{Csaki:2000zn}
\begin{equation}
\begin{aligned}
&ds^2 = e^{-2\sigma(y)}\left[1-2F(x^\mu,y)\right] \eta_{\mu\nu} dx^\mu dx^\nu + \left[1+2F(x^\mu,y)\right]^2dy^2 \:,\\
&\Phi_i(x^\mu,y) = \phi_i(y) + \varphi_i(x^\mu,y) \:.
\end{aligned}
\end{equation}
We work to linear order in the perturbations $F$ and $\varphi_i$. The equations for these perturbations consist of two of Einstein's equations and the Euler-Lagrange
equations:
\begin{align}
\label{eq:pert-flin-1}
&6M^3(F' - 2 \sigma' F) = \phi_i' \varphi_i \:,\\
\label{eq:pert-fquad-1}
&6M^3(-e^{2\sigma} \Box F - 2 \sigma' F' + F'') = 2 \phi_i' \varphi_i' + 2 \evbg{\lambda} F + \evbg{\lambda_i} \varphi_i \:,\\
\label{eq:pert-phi-1}
&e^{2\sigma} \Box \varphi_i + \varphi_i'' - 4 \sigma' \varphi_i' - 6 F' \phi_i'
    - \evbg{(4V_i + 2\lambda_i)} F
    - \evbg{(V_{ij} + \lambda_{ij})} \varphi_j = 0 \:.
\end{align}
Note that one of the redundant Einstein's equations has been
omitted and it can be easily obtained by using Eq.~\eqref{eq:pert-flin-1}
and Eq.~\eqref{eq:pert-fquad-1}.

We now go to conformal coordinates $z$ to eliminate the $e^{2\sigma}$
factor from the $\Box$'s, and rescale the fields by a factor of
$e^{3\sigma/2}$ to help eliminate first derivatives.  That is, we make the following change of variables and field redefinitions
\begin{equation}
\label{conformal}
\begin{aligned}
&dy = e^{-\sigma(y)} dz \:,\qquad
\sigma(y) = \sigma(z(y)) \:,\qquad
\phi_i(y) = \phi_i(z(y)) \:,\\
&F(y) = e^{\tfrac{3}{2}\sigma(z(y))} \tilde{F}(z) \:,\qquad
\varphi_i(y) = e^{\tfrac{3}{2}\sigma(z(y))} \tilde{\varphi}_i(z) \:.
\end{aligned}
\end{equation}
In terms of the conformal coordinate $z$ and the new fields
$\tilde{F}$ and $\tilde{\varphi}_i$, Eqs.~\eqref{eq:pert-flin-1} through~\eqref{eq:pert-phi-1} become, respectively,
\begin{align}
\label{eq:pert-flin-2}
&6M^3(-\tfrac{1}{2} \sigma' \tilde{F} + \tilde{F}') = \phi_i' \tilde\varphi_i \:,\\
\label{eq:pert-fquad-2}
&6M^3 \left[ -\Box \tilde{F} + \tilde{F}'' + 2 \sigma' \tilde{F}'
    + \left( \tfrac{3}{4} \sigma'^2 + \tfrac{3}{2} \sigma'' \right) \tilde{F} \right]
    = 2 \phi_i' \tilde\varphi_i'
        + 2 e^{-2\sigma} \evbg{\lambda} \tilde{F} \nonumber\\
&\qquad\qquad\qquad
        + \left( 3\sigma'\phi_i' + e^{-2\sigma}\evbg{\lambda_i} \right) \tilde\varphi_i \:,\\
\label{eq:pert-phi-2}
&\Box \tilde{\varphi}_i + \tilde{\varphi}_i''
    + \left( -\tfrac{9}{4} \sigma'^2 + \tfrac{3}{2} \sigma'' \right) \tilde\varphi_i
    - 3 \phi_i' \left( 3\sigma'\tilde{F} + 2\tilde{F}' \right) \nonumber\\
&\qquad\qquad\qquad
    - e^{-2\sigma} \evbg{(4V_i + 2\lambda_i)} \tilde{F}
    - e^{-2\sigma} \evbg{(V_{ij} + \lambda_{ij})} \tilde\varphi_j = 0 \:.
\end{align}
We can make judicious use of Eq.~\eqref{eq:pert-flin-2}, and its
derivative, to eliminate all first derivatives of $\tilde{F}$ and
$\tilde\varphi_i$ in the other two equations.  We also make use of
the background equation for $\phi_i$.  This allows us to obtain
\begin{align}
\label{eq:pert-flin-3}
&6M^3(\tilde{F}' - \tfrac{1}{2} \sigma' \tilde{F}) - \phi_i' \tilde\varphi_i = 0 \:,\\
\label{eq:pert-fquad-3}
&6M^3 \left[ \Box \tilde{F} + \tilde{F}''
    - \left( \tfrac{9}{4} \sigma'^2 + \tfrac{5}{2} \sigma'' \right) \tilde{F} \right]
    + 2 e^{-2\sigma} \evbg{\lambda} \tilde{F}
    + \left( -2\phi_i'' + e^{-2\sigma} \evbg{\lambda_i} \right) \tilde\varphi_i = 0 \:,\\
\label{eq:pert-phi-3}
&\Box \tilde{\varphi}_i + \tilde{\varphi}_i''
    + \left( -\tfrac{9}{4} \sigma'^2 + \tfrac{3}{2} \sigma'' \right) \tilde\varphi_i
    + \left( -4\phi_i'' + 2e^{-2\sigma}\evbg{\lambda_i} \right) \tilde{F} \nonumber\\
&\qquad\qquad\qquad\qquad\qquad
    - \frac{1}{M^3}\phi_i'\phi_j'\tilde{\varphi}_j
    - e^{-2\sigma} \evbg{(V_{ij} + \lambda_{ij})} \tilde\varphi_j = 0 \:.
\end{align}
This is a set of $N+2$ equations for $N+1$ functions of $z$.

\subsection{Simplifying the perturbation equations}

The main aim of this section is to manipulate and simplify the
above $N+2$ equations in order to obtain a set of coupled
Schr\"odinger-like equations, which can be solved to determine
the spin-0 spectrum.  Before simplifying the equations, we need
to understand precisely their redundancy.
Let $\alpha$, $\beta$ and $\gamma_i$ be equal to the
left-hand-sides of Eqs.~\eqref{eq:pert-flin-3} through~\eqref{eq:pert-phi-3}
respectively. Then the Euler-Lagrange and
Einstein's equations amount to
\begin{equation}
\label{eq:alpha-beta-gamma-system}
\alpha = \beta = \gamma_i = 0 \:.
\end{equation}
One can show that the following relation holds \emph{in the bulk}:
\begin{equation}
\label{eq:redundant}
\Box \alpha + \alpha''
    - \left( \tfrac{9}{4} \sigma'^2 + \tfrac{3}{2} \sigma'' \right) \alpha
    - \beta' + \tfrac{1}{2} \sigma' \beta + \phi_i' \gamma_i = 0 \:.
\end{equation}
Thus, there is a certain amount of redundancy in the $N+2$
equations.  The redundancy is quantified precisely
by Eq.~\eqref{eq:redundant}.  In particular, if we solve
$\alpha=\beta=0$ then we automatically have
$\phi_i'\gamma_i=0$ and solving for $N-1$ of the
$\gamma_i$'s is enough to solve for the system (that is, we have
eliminated one of the scalars $\tilde{\varphi}_i$).  Counting the number
of free integration constants (ICs), we have 2 for $\alpha=0$
and $\beta=0$ together, and $2(N-1)$ for $N-1$ of the $\gamma_i$'s.
This is a total of $2N$ free ICs.

We can make other choices for eliminating an equation.  For
example, if we choose to solve \emph{all} $\gamma_i=0$ and
$\beta=0$, then we will have an equation left over for $\alpha$:
\begin{equation}
\label{eq:alpha}
\Box \alpha + \alpha''
    - \left( \tfrac{9}{4} \sigma'^2 + \tfrac{3}{2} \sigma'' \right) \alpha = 0 \:.
\end{equation}
Solving $\beta=\gamma_i=0$ only
solves $\alpha$ such that it is a solution to the above
differential equation.  Counting the total
number of free ICs: 2 each from all $N$ of the $\gamma_i$'s and 2
from $\beta=0$. However, 2 of these ICs have to be used in order to pick the $\alpha=0$ solution of Eq.~\eqref{eq:alpha} and so we end up with a total of  $2N$ \emph{free} ICs for
the system, the same as in the alternative choice above.

We can also choose to solve $\alpha=0$ and $\gamma_i=0$
with $2N+1$ free ICs.  Then
Eq.~\eqref{eq:redundant} becomes $-\beta'+\tfrac{1}{2} \sigma' \beta=0$
and we need to have 1 constraining IC from demanding that
we pick out the $\beta=0$ solution from this differential equation.
Thus there are again only $2N$ free ICs for the system.

These arguments hold for solutions of the system in the bulk.
Without branes, the system is completely specified by $2N$
integration constants.  If there are branes in the set-up, the
situation is the same, as the values of fields on the brane are
continuous and hence the same as the values in the bulk just to
the sides of the brane(s).  Derivatives of the fields are discontinuous
on the brane, and hence undefined.  Instead, one has freedom to
choose the field derivatives on one side of the brane or the other,
and then uses the `jump' conditions to determine the derivative on
the opposite side.  The jump conditions are found by integrating
Eqs.~\eqref{eq:pert-fquad-3} and~\eqref{eq:pert-phi-3} over each
brane; see~\cite{Csaki:2000zn} for details.

We choose to solve the system $\beta=\gamma_i=0$, as it leaves us
with the most `symmetric' system of equations.  Solutions to this
particular set of equations will include all solutions of the
true system (by that we mean Eq.~\eqref{eq:alpha-beta-gamma-system}),
as well as some additional solutions which do not satisfy all the
boundary conditions of Einstein's equations.  For us, this is all
we shall require, as we are going to show that the extended set of
solutions does not contain certain modes (tachyonic and/or zero
modes), and so the full system cannot therefore contain such modes.

If one makes the definitions
\begin{align}\label{redefine}
\tilde\varphi_i(z) &= M^{3/2} \psi_i(z) \:,\\
\tilde{F}(z) &= \chi(z)/\sqrt{12} \:,
\end{align}
 then the equations
$\beta=\gamma_i=0$, Eqs.~\eqref{eq:pert-fquad-3}, and~\eqref{eq:pert-phi-3},
become
\begin{equation}
\label{eq:sym-spin0}
\begin{aligned}
&\Box \chi + \chi'' - (\mcv_{00} + \mcb_{00}) \chi - (\mcv_{0i} + \mcb_{0i}) \psi_i = 0 \:,\\
&\Box \psi_i + \psi_i'' - (\mcv_{ij} + \mcb_{ij}) \psi_j - (\mcv_{0i} + \mcb_{0i}) \chi = 0 \:,
\end{aligned}
\end{equation}
where
\begin{equation}
\begin{aligned}
\mcv_{00} &= \frac{9}{4} \sigma'^2 + \frac{5}{2} \sigma'' \:,\\
\mcv_{0i} &= \frac{2}{\sqrt{3M^3}} \phi_i'' \:,\\
\mcv_{ij} &= \left( \frac{9}{4} \sigma'^2 - \frac{3}{2} \sigma'' \right) \delta_{ij}
    + \frac{1}{M^3}\phi_i'\phi_j' + e^{-2\sigma} \evbg{V_{ij}} \:,
\end{aligned}
\end{equation}
and the brane terms are
\begin{equation}
\label{braneterms}
\begin{aligned}
\mcb_{00} &= \frac{1}{3M^3} e^{-2\sigma} \evbg{\lambda} \:,\\
\mcb_{0i} &= \frac{1}{\sqrt{3M^3}} e^{-2\sigma} \evbg{\lambda_i} \:,\\
\mcb_{ij} &= e^{-2\sigma} \evbg{\lambda_{ij}} \:.
\end{aligned}
\end{equation}
Note the symmetry of the cross-coupling in Eq.~\eqref{eq:sym-spin0}. We can use Eq.~\eqref{eq:sym-spin0} to solve for the physical spin-0
spectrum, which includes the eigenvalues and corresponding
extra-dimensional profiles.

A physical mode is defined as having the same $x^\mu$ dependence
in $\chi$ and $\psi_i$, but possibly different $z$ dependence.
Separation of variables then proceeds by defining
\begin{equation}
\begin{aligned}
\chi(x^\mu, z) = f_0(z) \eta(x^\mu) \:,\\
\psi_i(x^\mu, z) = f_i(z) \eta(x^\mu) \:,
\end{aligned}
\end{equation}
where $f_m(z)$, $m=0,1,\ldots,N$ are the extra-dimensional
profiles and $\eta(x^\mu)$ the 4D mode.  To find the spectrum of
mass states of $\eta$, we perform a Fourier transform on the
variable $x^\mu$: $\Box \eta \to E \eta$, where $E$ is the
eigenvalue, proportional to the mass squared of the $\eta$ mode.
Eq.~\eqref{eq:sym-spin0} now becomes a coupled eigenvalue
equation of the form 
\begin{equation}
\label{eq:schro}
H
\begin{pmatrix} f_0 \\ f_i \end{pmatrix}
=
\begin{pmatrix} -\partial_z^2 + \mcv_{00} + \mcb_{00} & \mcv_{0j} + \mcb_{0j} \\ \mcv_{0i} + \mcb_{0i} & -\partial_z^2 \delta_{ij} + \mcv_{ij} + \mcb_{ij} \end{pmatrix}
\begin{pmatrix} f_0 \\ f_j \end{pmatrix}
= E
\begin{pmatrix} f_0 \\ f_i \end{pmatrix} \:.
\end{equation}
In the matrix multiplication there is a sum over repeated $j$ indices
from 1 to $N$, and we shall use such a notation in subsequent equations.

This equation is a system of $N+1$
coupled Schr\"odinger equations, with a symmetric coupling
potential. The symmetry implies that the eigenvalues of the system
will be real, so long as certain boundary conditions are satisfied.
This equation is one of the main results of the paper.  It allows
one to determine the spectrum of physical spin-0 modes with $N$
scalars coupled to gravity.  It is valid for an arbitrary potential
$V$, with or without branes at the edges of the extra dimension, and
for generic topology of the extra dimension.

\subsection{The case without branes}
\label{sec:nobranes}

We are now going to specialize to the case where the brane terms
are absent; $\lambda=0$. A particular example of such models is one where a domain wall replaces the fundamental brane~\cite{Csaki:2000fc}. In this case we know exactly how the
perturbations behave at the boundaries, and can proceed to determine
the spectrum.  We shall show, using the fake supergravity
approach, that there are no tachyonic modes for these models.  We present conditions for the non-existence of tachyonic modes for models with fundamental branes in the following section.

In order to prove stability for a system with $N$ scalar fields and
the gravitational perturbation $\chi$, we need to show that the
eigenvalues $E$ are strictly positive for the given boundary conditions. Our analysis so far has been for a general potential $V$ and general
background configuration of scalar fields.  If we specialize to
configurations generated by the fake supergravity approach, we can
prove that the eigenvalues $E$ of this system are non-negative, and
in some cases strictly positive.

The key for proving such positivity is the observation that one can
write the perturbation potential as
\begin{equation}
\mcv
  = \begin{pmatrix} \mcv_{00} & \mcv_{0j} \\ \mcv_{0i} & \mcv_{ij} \end{pmatrix}
  = S^2 + S'  \:,
\end{equation}
where
\begin{equation}\label{generalS}
S = e^{-\sigma}
\left.
\begin{pmatrix}
\tfrac{1}{12 M^3} W &
\,\,\tfrac{1}{\sqrt{3 M^3}} W_{j} \\
\tfrac{1}{\sqrt{3 M^3}} W_{i} &
\,\,\tfrac{-1}{4 M^3}\delta_{ij} W
    + W_{ij}
\end{pmatrix}\right\vert_\text{bg}\,.
\end{equation}
Then, 
\begin{equation}\label{generalH}
H=(\partial_z + S^{\dagger})(-\partial_z + S)  \:.
\end{equation}
We have shown in Appendix~\ref{sec:hamiltonian_app} that the eigenvalues satisfy $E\ge0$ whenever 
\begin{equation}
\Psi^{\dagger}\left(-\partial_z + S \right) \Psi|_{\rm boundary} = 0\,,
\end{equation}
where $\Psi = (f_0,f_i)^T$.

As is obvious from the above equation, boundary conditions play a crucial role on the positivity of the eigenvalues and therefore on the stability of the system. Here we discuss the boundary conditions with general $N$ for the following
specific topologies: 

{\bf Full-interval space:}
On a full interval there are no restrictions, such as parity or 
Dirichlet or Neumann boundary conditions, on the perturbation wave 
function $\Psi$.  We only require that the perturbations are
normalizable over the full space.  The boundaries are at $z_1$ and
$z_2$, which can be at infinity or at finite values as in the case
of a soft-wall model, where space ends at a physical singularity. 
As seen in Eq.~\eqref{conformal} solutions in conformal coordinates $z$ are related to the ones in $y$ coordinates through the warp factor which diverges at large $z$. We see that in order to have normalizable solutions, the rescaled perturbations in conformal coordinates must vanish at the boundaries. Therefore, $\Psi^{\dagger}\left(-\partial_z + S \right) \Psi|_{z = z_{1,2}} = 0 $ is satisfied.

{\bf Half-interval orbifold space:}
Taking a full-interval space and identifying $y$ with $-y$ yields 
an orbifold, with effective boundaries at $z=0$ and $z=z_1$. 
Note that for such a topology there will be a non-dynamical brane at 
the origin, but this will not contribute to the brane terms $\mcb$ in 
the effective potential for the perturbations.  Furthermore, we do not 
need to appeal to string theory for information on the physics of this 
non-dynamical brane; it is just an orbifold fixed plane.
The boundary conditions for $\Psi$ at $z=0$ are now equivalent to a 
choice of parity for the warp factor and the scalars. 
The warp factor must be even in order to localize gravity. 
Scalar fields that are odd vanish at $z=0$, even scalars have 
$\partial_z\varphi_i$ vanishing at $z=0$. It is also easy to verify that $\Psi^{\dagger}S\,\Psi$ has odd parity, and therefore it also vanishes at $z=0$. As such, the condition  $\Psi^{\dagger}\left(-\partial_z + S \right) \Psi|_{z=0} = 0 $ is always satisfied.
For the $z=z_1$ boundary, the scalar perturbations vanish
as for the full-interval case discussed above due to the normalizable condition on the perturbations.

Having explicitly shown that the boundary terms vanish, which is expected for a Hermitian Hamiltonian corresponding to a physical problem, we can conclude that without
dynamical branes models with $N$ scalars 
coupled to gravity do not have any tachyonic modes. Note that our results so far apply to both of these scenarios while in the following sections for explicit examples with two or more scalar fields we will concentrate on the half-interval case.

\subsection{The case with fundamental branes}

For completeness we would also like to give sufficient conditions for the non-existence of tachyonic modes for models with $N$ scalars coupled to gravity in the presence of fundamental branes with $\lambda \ne 0$. The profiles for the scalar fields for models with fundamental branes satisfy the effective Schr\"odinger equation given by Eq.~\eqref{eq:schro} where the brane terms $\mcb_{mn}$ are given in Eq.~\eqref{braneterms}. Repeating the analysis of Appendix~\ref{sec:hamiltonian_app} with brane terms we find that the effective Schr\"odinger equation can be written as
\begin{eqnarray}
\int\,dz\,|\mathcal{S}\Psi|^2 + \int\,dz\,\partial_z\left(\Psi^{\dagger} \mathcal{S} \Psi\right) + \int\,dz\,\left(\Psi^{\dagger}\mcb\,\Psi\right) = E\int\,dz|\Psi|^2\,,
\end{eqnarray}
where $\mathcal{S}=(-\partial_z+S)$ with $S$ given by Eq.~\eqref{generalS} and the brane terms are
\begin{equation}
\mcb = \sum_{\alpha}e^{-2\sigma}
\begin{pmatrix}
\tfrac{1}{3 M^3} \lambda^{\alpha} &
\tfrac{1}{\sqrt{3 M^3}} \lambda_{j}^{\alpha} \\
\tfrac{1}{\sqrt{3 M^3}} \lambda_{i}^{\alpha} &
\lambda_{ij}^{\alpha}
\end{pmatrix} \:
\delta(z-z_{\alpha})\,,
\end{equation}
where the sum over $\alpha$ is over all fundamental branes in the model. For $E\ge 0$ the sufficient condition is then modified to be 
\begin{equation}
\Psi^{\dagger}\left(-\partial_z + S\right)\Psi|_{{\rm boundary}} + \Psi^{\dagger}\mcb\Psi \, \ge 0\,.
\end{equation}
Analyzing this requirement for different models with different brane potential terms $\lambda^{\alpha}$ is outside the scope of this work.

\section{Analyzing Zero-Modes}\label{sec:specificn}
As we have seen in the previous section, the eigenvalues of the spin-0
sector are guaranteed to be non-negative when constructing configurations
using the fake supergravity approach. We must still analyze the existence of 
zero mode solutions to show complete stability. In particular this will
ensure that the size of the extra dimension is stabilized.

In this section, using the formalism we have developed, we will first
show that for the $N=1$ case the eigensystem reduces to a single
Schr\"odinger-like equation, and we determine the mass gap
of the spin-0 spectrum. For configurations with $N\ge 2$ scalar fields,
analyzing the general properties of the mass spectrum is a very difficult
problem.  As such, we restrict our attention to orbifold spaces where
the fields have definite parity and discuss a criterion for the
possible existence, or lack, of zero modes.  Guided by this criterion,
we analyze the zero modes in two example models with specific
superpotentials.  We shall explicitly construct a model with $N=2$
scalar fields that does not have a zero mode and therefore is stable.

\subsection{The $N=1$ case}

In this section we will analyze the system with one scalar coupled to gravity and show that it is stable with positive mass eigenvalues. The analysis is valid for general topologies.  Stability for models with one scalar field and no branes have been analyzed previously in Refs.~\cite{DeWolfe:1999cp, DeWolfe:2000xi}. For models with a soft-wall and a fundamental brane stability has been proven in Ref.~\cite{Cabrer:2009we} and for RS models with two fundamental branes in Refs.~\cite{Goldberger:1999uk, Csaki:2000zn}. 

We start by analyzing  the Einstein constraint equation,
Eq.~\eqref{eq:pert-flin-3}, and the second-order equation for $\chi$
from Eq.~\eqref{eq:schro}.  They are, respectively,
\begin{align}
& -f_0' + S_{00} f_0 + S_{0i} f_i = 0 \:,\\
& -f_0'' + \mcv_{00} f_0 + \mcv_{0i} f_i = E f_0\ \:.
\end{align}
As before, repeated $i$, and later $j$, indices are to be summed.
Combining them in an obvious way yields
\begin{equation}
-f_0''
  + \frac{S_{0i}\mcv_{0i}}{S_{0j}S_{0j}} f_0'
  + \left( \mcv_{00} - \frac{S_{00}S_{0i}\mcv_{0i}}{S_{0j}S_{0j}} \right) f_0
  + \left( \mcv_{0i} - \frac{S_{0i}S_{0k}\mcv_{0k}}{S_{0j}S_{0j}} \right) f_i
  = E f_0 \:.
\end{equation}
We can eliminate the $f_0'$ term, and obtain a Schr\"odinger-like
equation, by defining
\begin{equation}
f_0 = \sqrt{S_{0i}S_{0i}} \, g \:.
\end{equation}
This gives 
\begin{equation}
\label{eq:no-zero-mode}
-g''
  + \left( A^2 - A' + S_{0i}S_{0i} \right) g
  + \frac{1}{\sqrt{S_{0i}S_{0i}}} B_i f_i = E g \:,
\end{equation}
where
\begin{align}
A &= \frac{S_{0i}S_{ij}S_{0j}}{S_{0k}S_{0k}} \:,\\
B_i &= \mcv_{0i} - \frac{S_{0i}S_{0k}\mcv_{0k}}{S_{0j}S_{0j}} \:.
\end{align}
Equation~\eqref{eq:no-zero-mode} can be used to prove the non-existence
of a zero mode for a theory with one scalar field. For $N=1$ we have:
\begin{align}
A &= S_{11} \:,\\
B_i &= 0 \:.
\end{align}
Following the arguments in Appendix~\ref{sec:hamiltonian_app}, we see
that for wave functions  $\Psi \ne 0$, the eigenvalues of this
system are non-negative since the boundary terms vanish as we have
discussed previously. Equation~\eqref{eq:no-zero-mode} simplifies to
\begin{equation}
\label{eq:ne1diff}
(\partial_z - A)(-\partial_z - A) g + S_{0i}S_{0i} g = E g \:.
\end{equation}
Now we will use the same trick as in the analysis of
Appendix~\ref{sec:hamiltonian_app} and multiply Eq.~\eqref{eq:ne1diff}
from the left by $g^*$ and integrate over the extra dimension
$z$. Up to a surface term that vanishes we have
\begin{equation}
\label{eq:ne1eval}
\int\, dz \, \left|\left(-\frac{d}{dz}-A\right) g \right|^2
  + \int\,dz\,|S_{01} g |^2
  = E\int\,dz\,\left\vert g \right\vert^2\,.
\end{equation}
Consider the existence of a zero mode, with $E=0$.
In order to satisfy Eq.~\eqref{eq:ne1eval} for a generic
superpotential $W$, and for the given boundary conditions
by which the surface terms vanish, both of the terms on the left-hand
side must vanish simultaneously.  For a generic $W$ this means
that the field $g$, and hence $f_0$, has to vanish.  This shows
that for nontrivial $f_0$ there does not exist any zero mode. 

We can push the analysis a little further and provide a lower bound on the mass gap to the
first spin-0 state.  Dropping one of the terms on the left-hand
side of Eq.~\eqref{eq:ne1eval} gives the following inequality:
\begin{align}
E\int\,dz\,\left\vert g \right\vert^2
  &\ge \int\,dz\,|S_{01} g |^2 \\
  &\ge \min[S_{01}^2] \int\,dz\,|g|^2 \:,
\end{align}
where $\min[S_{01}^2]$ is the minimum of $S_{01}^2$ over $z$.
Thus we obtain a bound on the mass of the first spin-0 state:
\begin{equation}
\label{eq:n1-zero-mode-bound}
E \ge \min[S_{01}^2] \:.
\end{equation}

\subsection{The $N\ge2$ case}

Set-ups with more than one scalar field can in general posses a
zero mode.  Nevertheless, we shall provide a simple criterion that
can be used to find models which do not have a zero mode.  From
now on we must specialize to half-interval orbifold spaces so that
the $N$ scalar fields have definite parity.  This enables us to
formulate the following criterion:\\\\
{\it Zero-mode criterion: For a system of definite parity with $N$ scalar fields that couple to gravity, the number of independent normalizable zero modes with $E=0$ is at most equal to the number of even fields.}\\\\
This criterion is essentially a statement about integration constants.
In the fake supergravity approach, the background configuration is
given by the solutions to the first order equations, Eq~\eqref{w}.
This is a system of linear differential equations for $N+1$ functions,
which require $N+1$ Dirichlet boundary conditions for a unique solution.
The value of $\sigma(0)$ can always be chosen to be $0$ since any other
constant shift can be obtained by a redefinition of the coordinates
$x^{\mu}$.  The restriction to orbifold models further eliminates
those integration constants associated with odd-parity scalars, since
their field value must also vanish at $y=0$.  Unique solutions to the
fake supergravity equations are then parametrized by the integration
constants of the even-parity fields.  The final point is that zero
modes move us continuously through this space of solutions, so there
cannot be more zero modes than the number of even fields.

For our above argument to hold, we must show that the zero modes do
indeed take us from one solution to the next.  Looking at Eq.~\eqref{a3}
it is easy to see that a solution with $E=0$ is satisfied when
$(-\partial_z+S)\Psi = 0$, that is
\begin{equation}
\label{zmsolutions}
\begin{aligned}
& -f_0' + S_{00} f_0 + S_{0i} f_i = 0 \:,\\
& -f_j' + S_{0j} f_0 + S_{ij} f_i = 0 \:,
\end{aligned}
\end{equation}
where the matrix elements $S_{mn}$ are given explicitly in Eq.~\eqref{generalS}.
Now suppose there are normalizable zero mode solutions $(f_0^{(0)},f_{i}^{(0)})$ which satisfy the above equations. Working in $y$ coordinates, we now add these zero mode solutions as perturbations to the background configuration to define the following new fields
\begin{equation}
\label{newbackground}
\begin{aligned}
\bar{\sigma}\left(\bar{y}\right)&=\sigma (y) + \epsilon\, \frac{e^{3/2\sigma}}{\sqrt{12}}\, f_0^{(0)}(y)\,,\\
\bar{\phi}_i\left(\bar{y}\right)&=\phi_i(y) + \epsilon\,e^{3/2\sigma}M^{3/2}\,f_{i}^{(0)}(y)\,,
\end{aligned}
\end{equation}
where the new coordinate $\bar{y}$ is expressed in terms of the
zero mode perturbations as
\begin{equation}
d\bar{y} = dy\left(1+2\epsilon\,\frac{e^{3/2\sigma}}{\sqrt{12}}\,f_0^{(0)}\right)\,.
\end{equation}
$\epsilon$ in the above equations is a small parameter that
parametrizes the small perturbations.

Solving for the derivatives of the zero mode solutions, $f_0^{(0)\,'}$ and $f_{i}^{(0)\,'}$, in terms of the zero mode solutions themselves, by using Eq.~\eqref{zmsolutions}, it is easy to show that the new fields defined in Eq.~\eqref{newbackground} satisfy
\begin{equation}
\label{newspequations}
\begin{aligned}
\frac{d}{d\bar{y}}\bar{\sigma}(\bar{y}) &= \frac{1}{6M^3}\,W\left(\{\bar{\phi}_i\}\right)\,,\\
\frac{d}{d\bar{y}}\bar{\phi}_i(\bar{y}) &= W_i\left(\{\bar{\phi}_i\}\right)\,,
\end{aligned}
\end{equation}
up to first order in the perturbation parameter $\epsilon$.
We see that these new fields are themselves background solutions with the {\it same} superpotential $W$. This demonstrates that the zero mode solutions for the system of $N$ scalar fields coupled to gravity translates from one background solution to another, taking us from one set of integration constants in the $y$ frame to another set of integration constants in the $\bar{y}$ frame.
Note that this result is valid up to first order in $\epsilon$, so the zero mode solutions have to be strictly much smaller than the background solutions, restricting our reasoning to normalizable zero modes.
Since any given zero mode is associated with continuously changing
a Dirichlet boundary condition, the number of physical zero modes
cannot be more than the number of even-parity scalars.  This completes
the proof of our criterion.

\subsection{Explicit Examples}

In this section we look at two specific domain wall models with $N=2$
scalar fields coupled to gravity.  We have already shown that the spin-0
spectrum has strictly non-negative mass eigenmodes, and proven a criterion
relating the existence of zero modes to the parities of the background
scalar profiles.  The two models to be presented will form explicit
realizations of this criterion: one has an even scalar profile and a zero
mode, the other all odd profiles and no normalizable zero mode.  Aside
from this difference, both models are qualitatively the same.  The field
$\Phi_1$ will play the role of a dilaton and has a background solution
which diverges at finite $y$, generating a physical singularity and
cutting off the extra dimension, effectively compactifying it.
This is known as a soft wall~\cite{Karch:2006pv, Batell:2008zm, Batell:2008me, Cabrer:2009we}.
The second field $\Phi_2$ takes the form of a kink, creating a domain-wall
whose purpose is to replace the positive tension brane in usual soft-wall
set-ups.  The models we consider are on a half-interval orbifold space with 
definite parity, and the domain-wall
sits at the origin, acting as an effective boundary of the extra
dimension.  A domain-wall soft-wall model is an appropriate name for
this type of set-up.  In this section we are concerned primarily in the
stability of such models, and do not discuss any other phenomenology.
Whether or not these models solve the hierarchy problem is an interesting
question which we intend to address in future work.

\subsection*{Example 1}
The first model we consider has the following superpotential
\begin{equation}\label{ex1}
W\left(\Phi_1,\Phi_2\right) = \left(a\,\Phi_2 - b\,\Phi_2^3\right)\,e^{\nu\Phi_1}\,,
\end{equation}
where $\Phi_1$ and $\Phi_2$ are the dilaton and the kink fields respectively. $a>0$, $b>0$ and $\nu$ are parameters in the model. We write the fields in terms of their background solutions and perturbations as
\begin{equation}
\label{decompose}
\begin{aligned}
\Phi_1(x,y) &= \phi_1(y) +\varphi_1(x,y)\,,\\
\Phi_2(x,y) &= \phi_2(y) +\varphi_2(x,y)\,.
\end{aligned}
\end{equation}
The background fields satisfy
\begin{equation}
\begin{aligned}
\frac{d}{dy}\phi_1 &= \nu\, \left(a\,\phi_2 - b\,\phi_2^3\right)\,e^{\nu\phi_1}\,,\\
\frac{d}{dy}\phi_2 &= \left(a-3b\,\phi_2^2\right)\,e^{\nu\phi_1}\,,\\
\frac{d}{dy}\sigma &= \frac{1}{6M^3}\,\left(a\,\phi_2 - b\,\phi_2^3\right)\,e^{\nu\phi_1}\,.
\end{aligned}
\end{equation}
We choose the background solutions such that $\phi_1$ has even and $\phi_2$ has odd parity, that is
\begin{eqnarray}
\frac{d}{dy}\phi_1|_{y=0} = 0\,,\qquad
\phi_2(0) = 0\,.
\end{eqnarray}
\FIGURE[t]{
\includegraphics[width=.6\textwidth]{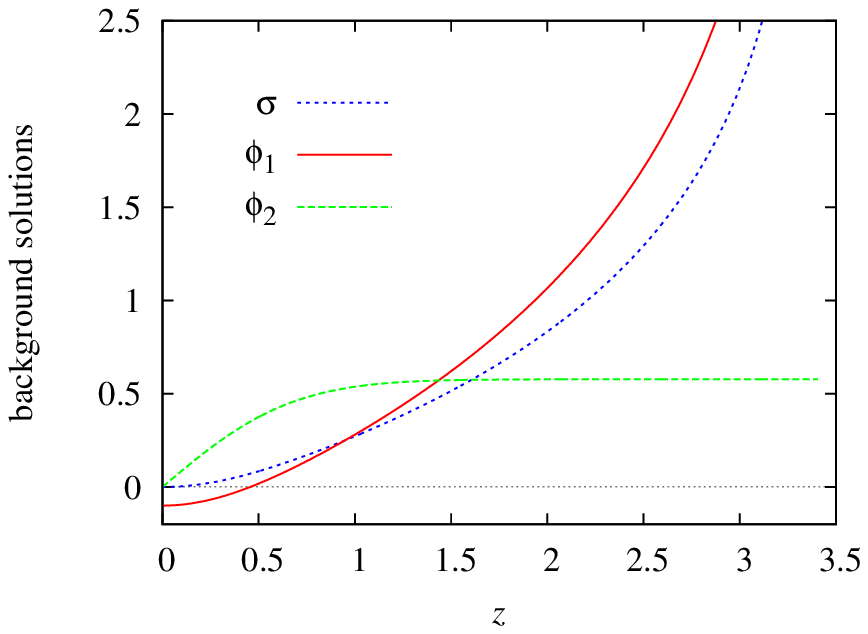}~
\caption{Numerical solutions for the backgrounds $\sigma$, $\phi_1$ and $\phi_2$ for example 1 with the parameter values $a=1$, $b=1$ and $\nu=1.4$. The free initial condition for $\phi_1(0)$ is chosen to be $-0.1$. The soft wall is at $z=3.41$ where $\sigma$ and $\phi_1$ diverge, ending the spacetime at a physical singularity. These solutions have definite parity and we have only plotted the $z\ge 0$ half.
\label{ex1fig}}
}

Using our criterion we know that this model will have at most one zero mode solution since we have one field whose background solution is even, which is the dilaton field. The nice thing about this simple superpotential choice is that the zero mode solution can be found analytically even though an analytic solution for the background fields is not possible. Fig.~\ref{ex1fig} shows the numerical solutions for the background fields for a particular parameter space point. 
Although we are not concerned here with the parameters of the model,
an important thing to mention is that our choice of $\nu=1.4$ allows
us to satisfy the equations of motion at the singularity; see
Ref~\cite{Cabrer:2009we} for details.\footnote{Note that, in addition
to $\Phi_i$, the parameters of our models are also dimensionful.  For
our plots we work with units where $6M^3=1$.}

Going to the conformal coordinates using Eq.~\eqref{conformal} and using the field redefinitions in Eq.~\eqref{redefine}, the zero mode  satisfies
\begin{equation}
\label{system}
\begin{aligned}
& -f_0^{(0)\,'} + S_{00} f_0^{(0)} + S_{01} f_{1}^{(0)} + S_{02}f_{2}^{(0)} = 0 \:,\\
& -f_{1}^{(0)\,'} + S_{01} f_0^{(0)} + S_{11} f_{1}^{(0)} + S_{12}f_{2}^{(0)} = 0 \:,\\
& -f_{2}^{(0)\,'} + S_{02} f_0^{(0)} + S_{12} f_{1}^{(0)} + S_{22}f_{2}^{(0)} = 0 \:,
\end{aligned}
\end{equation}
where the matrix elements $S_{mn}$ are given explicitly by Eq.~\eqref{generalS} using the superpotential of Eq.~\eqref{ex1}. It is easy to verify that the normalizable zero mode solution to the above system is given by
\begin{equation}
\begin{pmatrix} f_0^{(0)} \\ f_{1}^{(0)} \\ f_{2}^{(0)} \end{pmatrix} = 
\begin{pmatrix}  -N\,\frac{\nu}{\sqrt{2}}\,e^{-3/2\sigma}\\  N\,e^{-3/2\sigma} \\ 0\end{pmatrix}\,,
\end{equation}
where $N$ is a normalization constant. This zero mode physically corresponds to changes in the size of the extra dimension.

\subsection*{Example 2}
The second model we would like to analyze has the following superpotential
\begin{equation}\label{ex2}
W\left(\Phi_1,\Phi_2\right) = \alpha\,\sinh(\nu\,\Phi_1) + \left(a\,\Phi_2 - b\,\Phi_2^3\right)\,
\end{equation}
where $\Phi_1$ and $\Phi_2$ are again the dilaton and the kink fields. We write the fields in terms of their background and perturbations as in Eq.~\eqref{decompose}. As we have explained before the parity requirements on the gravitational background $\sigma$ forces us to have an odd parity superpotential. To satisfy this, we choose the background kink solution to have odd parity, as in the previous example, as well as the background dilaton solution:
\begin{eqnarray}
\phi_1(0) = 0\,,\qquad
\phi_2(0) = 0\,.
\end{eqnarray}
\FIGURE[t]{
\includegraphics[width=.6\textwidth]{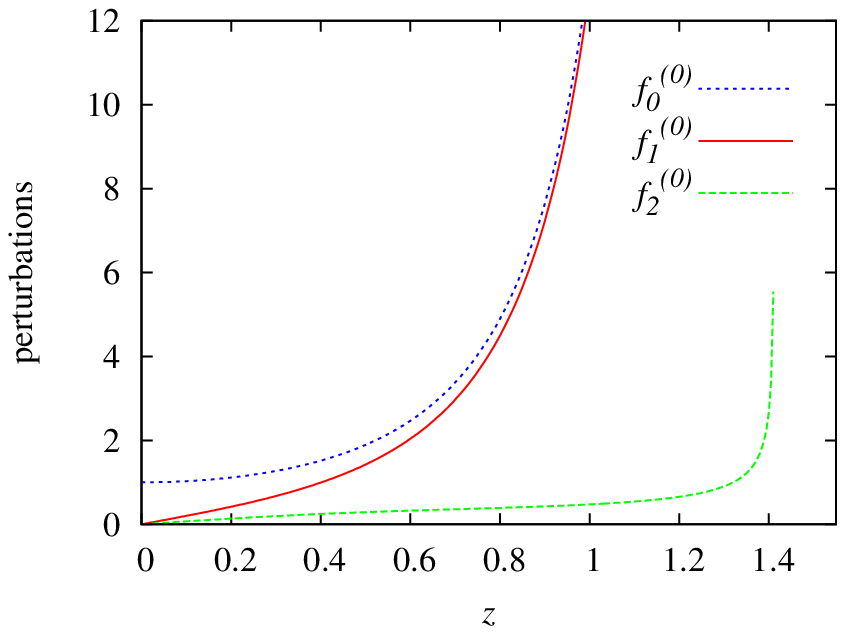}~
\caption{Numerical solutions for the zero mode solutions $f_0^{(0)}$, $f_{1}^{(0)}$ and $f_2^{(0)}$ for example 2 with the parameter values $a=0.5$, $b=0.3$, $\alpha=1$ and $\nu=1.4$. The free initial condition for $f_0(0)$ is chosen to be $1.0$. The figure shows that the zero mode solutions diverge in the region close to the soft wall, making the solution not normalizable. These solutions have definite parity and we have only plotted the $z\ge 0$ half.
\label{ex2fig}}
}
For this superpotential, the background solutions can be obtained analytically as
\begin{equation}
\begin{aligned}
\phi_1(y) &= \frac{2}{\nu}\,{\rm arctanh}\left(\tan\frac{\alpha \nu^2 y}{2}\right)\,,\\
\phi_2(y) &= \sqrt{\frac{a}{3b}}\,\tanh(\sqrt{3ab}\,y)\,.
\end{aligned}
\end{equation}
The position of the soft-wall singularity in $y$ coordinates can also be analytically determined to be
\begin{equation}
\label{eq:ys-value}
y_s= \frac{\pi}{2\alpha\nu^2}\,,
\end{equation}
which is the location in the extra dimension where the dilaton field and warp factor diverge.

The zero mode solution is given as a solution to $\mathcal{S}\Psi^{(0)} = 0$. The system of equations are given by Eqs.~\eqref{system}. We solve this system numerically using the superpotential in Eq.~\eqref{ex2}. Parities of the perturbations must be the same as the parities for the background fields:
\begin{eqnarray}
f_{0}^{(0)}(0) = 1\,,\qquad
f_{1}^{(0)}(0) = 0\,,\qquad
f_2^{(0)} (0) = 0\,,
\end{eqnarray}
where the scaling property of Eq.~\eqref{system} allows us to choose $f_0^{(0)}(0)=1$
without loss of generality.  If there exists a non-trivial zero mode
in this model, then it must possess these initial conditions.
Therefore, if these initial conditions lead to a non-normalizable
solution, the only normalizable solution is the trivial one.  We
find numerically that the solution diverges at the location of the
soft-wall, as shown in Fig.~\ref{ex2fig}, and the solutions are indeed not
normalizable.  This lack of a zero mode is consistent with our
criterion since, in this second example, all the scalars have odd parity.  
Furthermore, the existence of a non-normalizable solution to Eq.~\eqref{system} is
consistent with the proof of our criterion, which explicitly relied on 
the fact that perturbations must be small compared to the background
solution.

In summary, the model presented here does not contain a zero mode and
the physical size of the extra dimension is stabilized at the value
given by Eq.~\eqref{eq:ys-value}.

\section{Conclusions}\label{sec:conclusions}
Models with warped extra dimensions provide for interesting and rich
extensions beyond the Standard Model.  The classical backgrounds of
such models generally contain bulk scalars with non-trivial profiles.
Stability of the background is an important theoretical issue, in
particular, a model with a warped, compact extra dimension must
have the size of the extra dimension stabilized.  In this paper we
studied perturbative stability of $N$ scalars coupled to gravity by
analyzing the spin-2 and spin-0 spectra.  Although our results are
quite general, we paid particular attention to the case of domain-wall
models with a soft wall in an AdS$_5$ background.  These models
are interesting to study because, as we have shown, they provide a
purely field-theoretic mechanism for compactifying an extra dimension.

We have done part of our analysis using the fake supergravity
approach, which has been widely exploited to engineer analytically
tractable solutions to Einstein's equations.  This approach has also
been used to study stability of models where a single scalar field
with non-trivial profile is coupled to gravity in a warped
background~\cite{DeWolfe:1999cp, DeWolfe:2000xi, Freedman:2003ax}.
In this paper we have extended these previous studies to the case where
an arbitrary number of scalar fields couple minimally to gravity.

The first main result of this work is the derivation of Eq.~\eqref{eq:schro},
the coupled Schr\"odinger equation governing the spin-0 spectrum of
$N$ scalars coupled to gravity.  This equation is valid for an arbitrary
scalar potential, with or without additional brane terms.  Following
this, we specialized to potentials generated using fake supergravity
and presented our second main result: a system with $N$ scalar fields
coupled to gravity and without branes has no tachyonic modes in the
spin-0 sector.  This general result is valid for all types of models,
where the extra dimension may be infinite in size or finite with a
soft-wall, and where there may or may not be definite parity.  Extensions
to models with branes that have brane potential terms were also briefly
discussed.  This result generalizes previous studies on stability
of models with one scalar field coupled to
gravity~\cite{DeWolfe:1999cp, DeWolfe:2000xi}.  Using our formalism we
also studied the case with one scalar field and provided a lower bound
on the mass of the first spin-0 mode, Eq.~\eqref{eq:n1-zero-mode-bound}.

Our third main result is related to the existence of zero modes for
models with $N$ scalars coupled to gravity.  Zero mode solutions in
general destabilize the size of the extra dimension, and our aim was
to determine criteria which guaranteed the absence of such zero modes.
A general analytic study for models with an arbitrary number of scalar
fields is rather complex, and we restricted ourselves to scenarios
where the extra dimension has definite parity.  We have proven a
criterion that relates the number of zero modes to the parities of the 
scalars and used this result to show that zero modes are absent in
models where all background scalar profiles have odd parity.  We
demonstrated this by explicitly constructing two domain-wall models with
a soft wall, one of which admitted a zero mode and the other not.  The
latter is an example of a model that stabilizes a compact extra
dimension without using dynamical branes.  Whether these models are 
realistic models that can solve the hierarchy problem is a question to
be investigated in future work.

\acknowledgments
We would like to thank J.W. van Holten and M. Postma for useful
comments.
This research was supported by the Netherlands Foundation for
Fundamental Research of Matter (FOM) and the National Organization
for Scientific Research (NWO).

\appendix

\section{Analyzing the eigenvalues of the Hamiltonian}\label{sec:hamiltonian_app}
Take a system with an arbitrary number of scalars that satisfies
Schr\"odinger's equation, 
\begin{equation}\label{schro_app}
H \Psi =E \Psi \,.
\end{equation}
Below we show the standard result that if one can write $H$ as in supersymmetric quantum mechanics \cite{Cooper:1994eh},
\begin{equation}
H=(\partial_z + S)(-\partial_z + S) + K^{\dagger} K\,,
\end{equation} 
with Hermitian $S$, the eigenvalues of $H$ are non-negative for vanishing boundary terms. To see this we multiply Eq.~\eqref{schro_app} from left with $\Psi^{\dagger}$ and integrate over the extra dimension, which gives
\begin{eqnarray}\label{a3}
E \int\,dz\,|\Psi|^2 &=& \int\,dz\, \Psi^{\dagger} H  \Psi \,\nonumber\\
&=&\int\,dz\,\Psi^{\dagger}(\overrightarrow{\partial_z} + S^{\dagger})(-\overrightarrow{\partial_z} + S)\Psi +\int\,dz\,|K\,\Psi|^2\nonumber\\
&=&\int\,dz\,\Psi^{\dagger}(-\overleftarrow{\partial_z} + S^{\dagger})(-\overrightarrow{\partial_z} + S)\Psi +\int\,dz\,|K\,\Psi|^2 +\int\,dz\,\partial_z(\Psi^{\dagger}(-\partial_z + S)\Psi )\nonumber\\
&=&\int\,dz\,|\mathcal{S}\Psi|^2  +\int\,dz\,|K\,\Psi|^2 +\int\,dz\,\partial_z(\Psi^{\dagger}\mathcal{S}\Psi )\,,
\end{eqnarray}
where we defined 
\begin{equation}
\mathcal{S}\equiv (-\partial_z+S)\,.
\end{equation}
The arrows on the partial derivatives indicate which way they act. Notice that the last terms on the last two lines are the boundary terms. Since $|\mathcal{S}\Psi|^2\ge 0$ and $|K\Psi|^2\ge 0$, for an arbitrary $\Psi\ne 0$, we can immediately see that $E\ge 0$ if the boundary terms are zero, that is 
\begin{equation}
\Psi^{\dagger}\mathcal{S}\Psi|_{\rm boundary} = 0 \Rightarrow E\ge 0\,,
\end{equation}
which is satisfied when either $\Psi|_{\rm boundary}=0$ or $\mathcal{S}\Psi|_{\rm boundary} = 0$. 
In fact the requirement that a Hamiltonian is a self-adjoint operator for a physical problem already forces these boundary terms to vanish. We explicitly verify this for our particular models in Section~\ref{sec:nobranes}.
Note that these conclusions also hold when $K=0$.


\begin{thebibliography}{10}

\bibitem{Antoniadis:1990ew}
I.~Antoniadis, {\it {A Possible new dimension at a few TeV}},  {\em Phys.
  Lett.} {\bf B246} (1990) 377--384.

\bibitem{ArkaniHamed:1998rs}
N.~Arkani-Hamed, S.~Dimopoulos, and G.~R. Dvali, {\it {The hierarchy problem
  and new dimensions at a millimeter}},  {\em Phys. Lett.} {\bf B429} (1998)
  263--272, [\href{http://xxx.lanl.gov/abs/hep-ph/9803315}{{\tt
  hep-ph/9803315}}].

\bibitem{Antoniadis:1998ig}
I.~Antoniadis, N.~Arkani-Hamed, S.~Dimopoulos, and G.~R. Dvali, {\it {New
  dimensions at a millimeter to a Fermi and superstrings at a TeV}},  {\em
  Phys. Lett.} {\bf B436} (1998) 257--263,
  [\href{http://xxx.lanl.gov/abs/hep-ph/9804398}{{\tt hep-ph/9804398}}].

\bibitem{Randall:1999ee}
L.~Randall and R.~Sundrum, {\it {A large mass hierarchy from a small extra
  dimension}},  {\em Phys. Rev. Lett.} {\bf 83} (1999) 3370--3373,
  [\href{http://xxx.lanl.gov/abs/hep-ph/9905221}{{\tt hep-ph/9905221}}].

\bibitem{Randall:1999vf}
L.~Randall and R.~Sundrum, {\it {An alternative to compactification}},  {\em
  Phys. Rev. Lett.} {\bf 83} (1999) 4690--4693,
  [\href{http://xxx.lanl.gov/abs/hep-th/9906064}{{\tt hep-th/9906064}}].

\bibitem{Rubakov:1983bb}
V.~A. Rubakov and M.~E. Shaposhnikov, {\it {Do We Live Inside a Domain Wall?}},
   {\em Phys. Lett.} {\bf B125} (1983) 136--138.

\bibitem{Kehagias:2000au}
A.~Kehagias and K.~Tamvakis, {\it {Localized gravitons, gauge bosons and chiral
  fermions in smooth spaces generated by a bounce}},  {\em Phys. Lett.} {\bf
  B504} (2001) 38--46, [\href{http://xxx.lanl.gov/abs/hep-th/0010112}{{\tt
  hep-th/0010112}}].

\bibitem{Davies:2007xr}
R.~Davies, D.~P. George, and R.~R. Volkas, {\it {The standard model on a
  domain-wall brane?}},  {\em Phys. Rev.} {\bf D77} (2008) 124038,
  [\href{http://xxx.lanl.gov/abs/0705.1584}{{\tt arXiv:0705.1584}}].

\bibitem{Csaki:2000fc}
C.~Csaki, J.~Erlich, T.~J. Hollowood, and Y.~Shirman, {\it {Universal aspects
  of gravity localized on thick branes}},  {\em Nucl. Phys.} {\bf B581} (2000)
  309--338, [\href{http://xxx.lanl.gov/abs/hep-th/0001033}{{\tt
  hep-th/0001033}}].

\bibitem{Karch:2006pv}
A.~Karch, E.~Katz, D.~T. Son, and M.~A. Stephanov, {\it {Linear Confinement and
  AdS/QCD}},  {\em Phys. Rev.} {\bf D74} (2006) 015005,
  [\href{http://xxx.lanl.gov/abs/hep-ph/0602229}{{\tt hep-ph/0602229}}].

\bibitem{Batell:2008zm}
B.~Batell and T.~Gherghetta, {\it {Dynamical Soft-Wall AdS/QCD}},  {\em Phys.
  Rev.} {\bf D78} (2008) 026002, [\href{http://xxx.lanl.gov/abs/0801.4383}{{\tt
  arXiv:0801.4383}}].

\bibitem{Falkowski:2008fz}
A.~Falkowski and M.~Perez-Victoria, {\it {Electroweak Breaking on a Soft
  Wall}},  {\em JHEP} {\bf 12} (2008) 107,
  [\href{http://xxx.lanl.gov/abs/0806.1737}{{\tt arXiv:0806.1737}}].

\bibitem{Batell:2008me}
B.~Batell, T.~Gherghetta, and D.~Sword, {\it {The Soft-Wall Standard Model}},
  {\em Phys. Rev.} {\bf D78} (2008) 116011,
  [\href{http://xxx.lanl.gov/abs/0808.3977}{{\tt arXiv:0808.3977}}].

\bibitem{Delgado:2009xb}
A.~Delgado and D.~Diego, {\it {Fermion Mass Hierarchy from the Soft Wall}},
  {\em Phys. Rev.} {\bf D80} (2009) 024030,
  [\href{http://xxx.lanl.gov/abs/0905.1095}{{\tt arXiv:0905.1095}}].

\bibitem{MertAybat:2009mk}
S.~Mert~Aybat and J.~Santiago, {\it {Bulk Fermions in Warped Models with a Soft
  Wall}},  {\em Phys. Rev.} {\bf D80} (2009) 035005,
  [\href{http://xxx.lanl.gov/abs/0905.3032}{{\tt arXiv:0905.3032}}].

\bibitem{Gherghetta:2009qs}
T.~Gherghetta and D.~Sword, {\it {Fermion Flavor in Soft-Wall AdS}},  {\em
  Phys. Rev.} {\bf D80} (2009) 065015,
  [\href{http://xxx.lanl.gov/abs/0907.3523}{{\tt arXiv:0907.3523}}].

\bibitem{Cabrer:2009we}
J.~A. Cabrer, G.~von Gersdorff, and M.~Quiros, {\it {Soft-Wall Stabilization}},
   \href{http://xxx.lanl.gov/abs/0907.5361}{{\tt arXiv:0907.5361}}.

\bibitem{vonGersdorff:2010ht}
G.~von Gersdorff, {\it {From Soft Walls to Infrared Branes}},
  \href{http://xxx.lanl.gov/abs/1005.5134}{{\tt arXiv:1005.5134}}.

\bibitem{Cacciapaglia:2008ns}
G.~Cacciapaglia, G.~Marandella, and J.~Terning, {\it {The AdS/CFT/Unparticle
  Correspondence}},  {\em JHEP} {\bf 02} (2009) 049,
  [\href{http://xxx.lanl.gov/abs/0804.0424}{{\tt arXiv:0804.0424}}].

\bibitem{Falkowski:2008yr}
A.~Falkowski and M.~Perez-Victoria, {\it {Holographic Unhiggs}},  {\em Phys.
  Rev.} {\bf D79} (2009) 035005, [\href{http://xxx.lanl.gov/abs/0810.4940}{{\tt
  arXiv:0810.4940}}].

\bibitem{Goldberger:1999uk}
W.~D. Goldberger and M.~B. Wise, {\it {Modulus stabilization with bulk
  fields}},  {\em Phys. Rev. Lett.} {\bf 83} (1999) 4922--4925,
  [\href{http://xxx.lanl.gov/abs/hep-ph/9907447}{{\tt hep-ph/9907447}}].

\bibitem{Csaki:2000zn}
C.~Csaki, M.~L. Graesser, and G.~D. Kribs, {\it {Radion dynamics and
  electroweak physics}},  {\em Phys. Rev.} {\bf D63} (2001) 065002,
  [\href{http://xxx.lanl.gov/abs/hep-th/0008151}{{\tt hep-th/0008151}}].

\bibitem{Toharia:2007xe}
M.~Toharia and M.~Trodden, {\it {Metastable Kinks in the Orbifold}},  {\em
  Phys. Rev. Lett.} {\bf 100} (2008) 041602,
  [\href{http://xxx.lanl.gov/abs/0708.4005}{{\tt arXiv:0708.4005}}].

\bibitem{Toharia:2007xf}
M.~Toharia and M.~Trodden, {\it {Existence and Stability of Non-Trivial Scalar
  Field Configurations in Orbifolded Extra Dimensions}},  {\em Phys. Rev.} {\bf
  D77} (2008) 025029, [\href{http://xxx.lanl.gov/abs/0708.4008}{{\tt
  arXiv:0708.4008}}].

\bibitem{Kobayashi:2001jd}
S.~Kobayashi, K.~Koyama, and J.~Soda, {\it {Thick brane worlds and their
  stability}},  {\em Phys. Rev.} {\bf D65} (2002) 064014,
  [\href{http://xxx.lanl.gov/abs/hep-th/0107025}{{\tt hep-th/0107025}}].

\bibitem{Toharia:2008ug}
M.~Toharia, {\it {Odd Tachyons in Compact Extra Dimensions}},
  \href{http://xxx.lanl.gov/abs/0803.2503}{{\tt arXiv:0803.2503}}.

\bibitem{Toharia:2010ex}
M.~Toharia, M.~Trodden, and E.~J. West, {\it {Scalar Kinks in Warped Extra
  Dimensions}},  \href{http://xxx.lanl.gov/abs/1002.0011}{{\tt
  arXiv:1002.0011}}.

\bibitem{DeWolfe:1999cp}
O.~DeWolfe, D.~Z. Freedman, S.~S. Gubser, and A.~Karch, {\it {Modeling the
  fifth dimension with scalars and gravity}},  {\em Phys. Rev.} {\bf D62}
  (2000) 046008, [\href{http://xxx.lanl.gov/abs/hep-th/9909134}{{\tt
  hep-th/9909134}}].

\bibitem{Freedman:2003ax}
D.~Z. Freedman, C.~Nunez, M.~Schnabl, and K.~Skenderis, {\it {Fake Supergravity
  and Domain Wall Stability}},  {\em Phys. Rev.} {\bf D69} (2004) 104027,
  [\href{http://xxx.lanl.gov/abs/hep-th/0312055}{{\tt hep-th/0312055}}].

\bibitem{DeWolfe:2000xi}
O.~DeWolfe and D.~Z. Freedman, {\it {Notes on fluctuations and correlation
  functions in holographic renormalization group flows}},
  \href{http://xxx.lanl.gov/abs/hep-th/0002226}{{\tt hep-th/0002226}}.

\bibitem{Bazeia:2010yp}
D.~Bazeia, M.~M. Ferreira, Jr., A.~R. Gomes, and R.~Menezes, {\it
  {Lorentz-violating effects on topological defects generated by two real
  scalar fields}},  {\em Physica} {\bf D239} (2010) 942--947,
  [\href{http://xxx.lanl.gov/abs/1001.5286}{{\tt arXiv:1001.5286}}].

\bibitem{Cooper:1994eh}
F.~Cooper, A.~Khare, and U.~Sukhatme, {\it {Supersymmetry and quantum
  mechanics}},  {\em Phys. Rept.} {\bf 251} (1995) 267--385,
  [\href{http://xxx.lanl.gov/abs/hep-th/9405029}{{\tt hep-th/9405029}}].

\end{thebibliography}

\providecommand{\href}[2]{#2}\begingroup\raggedright\endgroup

\end{document}